\documentclass{article}

\usepackage{arxiv}

\usepackage{ulem}
\usepackage[utf8]{inputenc} 
\usepackage[T1]{fontenc}    
\usepackage{hyperref}       
\usepackage{url}            
\usepackage{booktabs}       
\usepackage{amsmath, amsfonts}       
\usepackage{nicefrac}       
\usepackage{microtype}      
\usepackage{lipsum}		
\usepackage{graphicx}
\usepackage[numbers,sort&compress,square]{natbib}
\usepackage{doi}

\usepackage{multirow}
\usepackage{lineno}
\usepackage{soul}
\usepackage{comment}

\graphicspath{{./Figures-png/}, {./Figures-png/power3}}

\usepackage{xcolor}

\newcommand{\gr}[1]{\textcolor{lightgray}{#1}}

\title{Assessing Emulator Design and Training for Modal Aerosol Microphysics Parameterizations in E3SMv2}

\author{ 
    \href{https://orcid.org/0000-0001-5548-0265}{\includegraphics[scale=0.06]{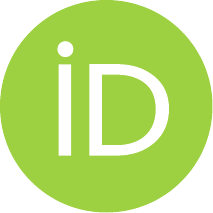}
    \hspace{1mm}Shady E. Ahmed} \\
	Advanced Computing, Mathematics, and Data Division \\
	Pacific Northwest National Laboratory\\
	Richland, WA 99354 \\
	\texttt{shady.ahmed@pnnl.gov} \\
	\And
	\href{https://orcid.org/0000-0001-5294-4116}{\includegraphics[scale=0.06]{orcid.pdf}\hspace{1mm}Hui Wan} \\
	Earth and Biological Sciences Directorate\\
    Pacific Northwest National Laboratory\\
	Richland, WA 99354 \\
    \And
    \href{https://orcid.org/0000-0002-0829-3907}{\includegraphics[scale=0.06]{orcid.pdf}
    \hspace{1mm}Saad Qadeer} \\
	Advanced Computing, Mathematics, and Data Division \\
	Pacific Northwest National Laboratory\\
	Richland, WA 99354 \\
    \And
    \href{https://orcid.org/0000-0002-9928-5637}{\includegraphics[scale=0.06]{orcid.pdf}
    \hspace{1mm}Panos Stinis} \\
	Advanced Computing, Mathematics, and Data Division \\
	Pacific Northwest National Laboratory\\
	Richland, WA 99354 \\
	\And
	  Kezhen Chong \\
	Earth and Biological Sciences Directorate\\
    Pacific Northwest National Laboratory\\
    Now at: Equifax Inc., Atlanta, GA\\
	\And
	  Mohammad Taufiq Hassan Mozumder\\
	Earth and Biological Sciences Directorate\\
    Pacific Northwest National Laboratory\\
    Now at: California Air Resources Board, CA
    \And
	\href{https://orcid.org/0000-0003-0457-6368}{\includegraphics[scale=0.06]{orcid.pdf}\hspace{1mm}Kai Zhang} \\
	Earth and Biological Sciences Directorate\\
    Pacific Northwest National Laboratory\\
	Richland, WA 99354 \\
    \And
    \href{https://orcid.org/0000-0003-2103-312X}{\includegraphics[scale=0.06]{orcid.pdf}\hspace{1mm}Ann S. Almgren} \\
    Applied Mathematics Department \\
    Lawrence Berkeley National Laboratory \\
    Berkeley, CA 94720 \\
}

\renewcommand{\shorttitle}{ML Emulation of Aerosol Microphysics in E3SMv2}

\hypersetup{
pdftitle={ML Emulation of Aerosol Microphysics in E3SMv2},
pdfauthor={Shady E. Ahmed, Hui Wan, Saad Qadeer, Panos Stinis, Kezhen Chong, Mohammad Taufiq Hassan Mozumder, Kai Zhang, Ann S. Almgren},
pdfkeywords={Aerosol microphysics, Machine learning, E3SMv2, Emulation},
}

\begin{document}
\maketitle

\begin{abstract}
Toward the goal of using Scientific Machine Learning (SciML) emulators to improve the numerical representation of aerosol processes in global atmospheric models, we explore the emulation of aerosol microphysics processes under cloud-free conditions in the 4-mode Modal Aerosol Module (MAM4) within the Energy Exascale Earth System Model version 2 (E3SMv2). To develop an in-depth understanding of the challenges and opportunities in applying SciML to aerosol processes, we begin with a simple feedforward neural network architecture that has been used in earlier studies, but we systematically examine key emulator design choices, including architecture complexity and variable normalization, while closely monitoring training convergence behavior.

Our results show that optimization convergence, scaling strategy, and network complexity strongly influence emulation accuracy. When effective scaling is applied and convergence is achieved, the relatively simple architecture, used together with a moderate network size, can reproduce key features of the microphysics-induced aerosol concentration changes with promising accuracy. These findings provide practical clues for the next stages of emulator development; they also provide general insights that are likely applicable to the emulation of other aerosol processes, as well as other atmospheric physics involving multi-scale variability.
\end{abstract}


\newpage

\section{Introduction} \label{sec:intro}
With the rapid advancement and growing adoption of artificial intelligence (AI) and Scientific Machine Learning (SciML), the Earth system science community has increasingly sought to incorporate these methods into modeling, prediction, and analysis workflows. Recent progress (see, e.g., summary in \cite{eyring2024pushing}) has included emulation of global atmospheric and ocean models \cite{watt2023ace,duncan2024application,kochkov2024neural,duncan2025samudrace}, development of AI- and SciML-based cloud and turbulence parameterizations \cite{rasp2018deep,gentine2018could,beucler2023machine,han2020moist,han2023ensemble}, and the release of curated community datasets \cite{watson2022climatebench,yu2023climsim}.
Within this broader movement, the Energy Exascale Earth System Model (E3SM) community has identified several promising avenues for integrating AI and SciML in Earth system model (ESM) development and application, including full-model emulation, process emulation, parameter estimation (calibration), workflow automation, and model initialization or spin-up \cite{e3sm_website}. Some efforts have been directed toward applying AI and SciML to improve the representation of aerosol life cycles and their interactions with clouds and radiation \cite{geiss2025NeuralMie,silva2021activation,zheng2021estimating,zheng2021quantifying},
while many avenues remain to be explored.

In the realm of traditional numerical models based on partial differential equations (PDEs) and ordinary differential equations (ODEs), various global atmospheric models and chemical transport models have incorporated parameterizations of aerosol microphysics to simulate changes in the size and composition of aerosol particles caused by processes such as nucleation, condensation--evaporation, coagulation, and aging \cite{bauer2008matrix,liu2012mam3,zhang2012global,wang2020aerosols}. Due to the complexity of these processes as well as the sharp contrast between their characteristic scales and the typical grid spacing in global models, it has been challenging to develop computational methods that are both accurate and computationally efficient.
The modal method, which uses assumed classes of functions to represent the statistical distribution of particle size and composition (see, e.g., \cite{whitby1991modal,vignati2004m7,easter2004mirage}), offers a good trade-off between accuracy and cost, although current implementations still contain ample room for improvement \cite{zhang2010tropospheric,mann2014intercomparison}. ML-trained emulators are worth exploring for the purpose of addressing this challenge by providing fast and accurate emulators of the underlying physics.

Several recent studies have taken initial steps toward this goal. \citet{harder2021emulating} introduced one of the first ML emulators for aerosol microphysics, targeting the M7 parameterization suite \cite{vignati2004m7} in the ECHAM-HAM model \cite{stier2005aerosol,zhang2012global}, where they also explored the role of physical constraints \cite{harder2022physics}. More recently, \citet{bai2025data} developed a deep operator network for emulating the aerosol microphysics parameterization suite in E3SM version 2 (E3SMv2). Collectively, these studies reported a notably wide range for the coefficient of determination ($R^2$)---from roughly 0.48 to 0.99 \cite{harder2021emulating,harder2022physics,bai2025data}---revealing substantial variability in emulator performance. So far, the published studies lack a systematic analysis of why such variability arises and how specific architectural choices, training strategies, or properties of the problem being emulated contribute to these differences. In other words, the community has produced promising results but does not yet have a clear baseline or design-space understanding for aerosol microphysics emulation.
Such a baseline and understanding will provide a necessary foundation for future work, especially since the physical processes of interest present unique challenges for ML emulation, including, e.g., the wide dynamic ranges in aerosol properties and the physical constraints that must be preserved.

This study aims at addressing this gap in understanding by performing a controlled, detailed investigation of the design decisions that affect the accuracy and reliability of aerosol microphysics emulators. To disentangle emulator-dependent effects from complexities resulting from different modes of variability or boundary conditions in the Earth system, we start by restricting our analysis to one representative condition, namely boreal winter, and we defer it to future studies to investigate the topic of generalizability with respect to external factors such as seasonality, diurnal cycle, and aerosol source scenarios. 
This enables a clean evaluation of how model architecture, input/output configuration, and training setup affect emulator performance. In the first step, we focus on assessing what a standard feedforward neural network (FNN) can achieve in this controlled setting, so as to establish a clear, easy-to-obtain baseline for future developments.
We aim at emulating the clear-air (i.e., cloud-free) aerosol microphysics suite in E3SMv2 exercised in the default 4-mode configuration of the aerosol module, in simulations conducted at the current workhorse resolution of about 165~km horizontal grid spacing, so that our findings can directly inform E3SM-focused AI/ML development.
Nevertheless, various strategies presented in this paper, including the assessment of training convergence and the use of variable transformations to handle physical quantities spanning wide ranges of scales,
are generally applicable to the emulation of other parameterizations and PDE/ODE models.

While potential long-term plans include exploring AI/ML emulators to improve upon the current PDE-based representation of aerosol life cycles in E3SM by emulating, for example, measurements data and simulation output from much more detailed, physics-based, process-resolved numerical models, we begin by emulating the aerosol microphysics parameterization suite currently used by E3SMv2 so as to focus on emulator accuracy. This framing allows us to treat the problem as a controlled benchmark: the underlying parameterization is fixed, and the primary question becomes how accurately and robustly an ML emulator can reproduce its behavior. Emulating a parameterization suite currently in use also has immediate practical value, as microphysics can be a computationally expensive component of atmospheric aerosol models, and this was a key factor that motivated some of the earlier studies, e.g., \cite{harder2021emulating,harder2022physics}.

The remainder of the manuscript is organized as follows. 
Section~\ref{sec:method} introduces the aerosol microphysics module in E3SMv2, the input-output mapping to be emulated, and the neural network architecture used to learn that mapping. Section~\ref{sec:ml-results} describes the construction and training of the emulator, explores the design space and convergence behavior, and evaluates performance across target variables. A summary of the key results are presented in Section~\ref{sec:conclusion}.

\section{Methods} \label{sec:method}

\subsection{Aerosol microphysics in E3SMv2} \label{sub:amicphys}
The aerosol microphysics parameterization suite in E3SMv2 is a part of the Modal Aerosol Module (MAM) designed for global aerosol modeling \cite{wang2020aerosol,liu2016mam4,liu2012mam3,easter2004mirage}, which groups aerosol particles into lognormally distributed size modes. The abundance of particle number and mass in each mode is quantitatively described by the particle number mixing ratio of that mode and the mass mixing ratios of multiple chemical components. E3SMv2 uses the four-mode configuration, MAM4 \cite{wang2020aerosol,liu2016mam4}, which consists of Accumulation mode, Aitken mode, coarse mode, and primary carbon mode with seven chemical components: sulfate (so4), secondary organic aerosols (soa), black carbon (bc), primary organic matter (pom), sea salt (ncl), marine organic matter (mom), and mineral dust (dst).
Aerosol particles are also classified by their attachment state: \textit{interstitial} aerosols are found outside cloud droplets, whereas \textit{cloud-borne} aerosols reside within cloud droplets. The explicit representation of cloud-borne aerosols using separate prognostic equations \cite{ghan2006cloudborne,easter2004mirage} is a unique feature of MAM compared to many other aerosol models (see, e.g., \cite{stier2005aerosol,zhang2012global,mann2010description}); in addition, the assumptions about aerosol processes in partially or fully cloudy grid boxes can also differ significantly between MAM and other aerosol modules depending on the process in question. In order to allow comparison of our results with those from, e.g., \cite{harder2022physics}, and to make our conclusions potentially informative for other aerosol models, this study focuses on {\it interstitial} aerosol microphysics under {\it cloud-free} conditions; work on cloudy conditions and cloud-borne aerosols is deferred to the next step.

Interstitial aerosols correspond to 25 mixing ratios predicted in each grid cell of E3SMv2's atmospheric grid, see Table~\ref{tab:tracers}. These mixing ratios are influenced by physical processes including not only microphysics but also emissions, transport by resolved and unresolved air motions, gravitational settling, as well as interactions with clouds, rain, and the Earth's surface. The parameterization suite of aerosol microphysics is formulated as a set of coupled ODEs that affect 18 of the 25 interstitial aerosol mixing ratios (Table~\ref{tab:tracers}). In addition, 2 gas components, namely sulfuric acid gas (H2SO4) and the precursor of secondary organic aerosols (SOAG), exchange mass with interstitial aerosols through condensation and evaporation. This gives a total of 20 ($= 18 + 2$) mixing ratios that are affected by aerosol microphysics in cloud-free conditions (Table~\ref{tab:tracers}).

For the discrete time integration in E3SMv2 (and also in its predecessors, E3SMv1 \cite{wang2020aerosol} and CESM1 \cite{liu2016mam4,liu2012mam3}), aerosol microphysics is handled using operator splitting. Taking the workhorse resolution of E3SMv2 as an example, the atmosphere component has horizontal grid spacings of about 165~km and a main timestep of 30~min. Most of the parameterizations inside the atmosphere component are calculated once every 30~min unless sub-cycling or super-cycling is employed. Within the 30~min timestep, most parameterizations are calculated sequentially, letting one process (or process group) update the prognostic variables before passing the updated values to the next process (or group). Similarly, the aerosol microphysics code for cloud-free conditions takes as input the 20 mixing ratios described above as the initial conditions of a 30~min timestep; the ODEs are advanced for one timestep, and the new mixing ratios at the end of the timestep are passed back to the host model. The right-hand side of the ODEs represents only aerosol microphysics and no other processes are involved. The only exception is that the chemical production rate of the H2SO4 gas calculated by the chemistry module (which is a separate part of the E3SM atmosphere model) is included as a forcing term that remains constant within the 30~min. Earlier studies have shown that including this forcing term is important for the time-stepping accuracy of the microphysics ODEs \cite{wang2020aerosol,wan2013h2so4}.
In addition to the 20 input mixing ratios and the chemical production rate of the H2SO4 gas, integrating the microphysics ODEs also requires information of atmospheric conditions like temperature, pressure, and humidity, as well as aerosol properties such as the geometric mean dry and wet particle diameters and density. These are listed in Table~\ref{tab:other_input}.

\begin{table}[t!]
    \centering
    \caption{Interstitial aerosol mass, interstitial aerosol particle number, and gas precursor mass for which the concentrations (mixing ratios) are predicted by the MAM4 aerosol module in E3SMv2.
    The mixing ratios shown in light gray are not affected by the microphysics parameterizations in MAM4, and they do not affect the changes of mixing ratios shown in black; hence they are not included in ML emulation.
    The chemical components are: sulfate (so4), seconary organic aerosol (soa), black carbon (bc), primary organic matter (pom), sea salt (ncl), marine organic matter (mom), dust (dst), sulfuric acid gas (H2SO4), and the gas precursor of secondary organic aerosols (SOAG). Particle number mixing ratios are denoted by ``num". The labels ``a1'' through ``a4'' are labels of the lognormal modes in MAM4.
    } 
    \vspace{1pt}
    \renewcommand{\arraystretch}{1.2} 
    \scalebox{0.9}{
    \begin{tabular}{cccc|c}
    \hline
    \multicolumn{4}{c|}{{\bf Aerosols}}                            & {\bf Gas} \\
    \cline{1-4}
    {\bf Accumulation mode (a1)}  & {\bf Aitken mode (a2)}         &
    {\bf Coarse mode (a3)}        & {\bf Primary carbon mode (a4)} & {\bf Precursors} \\
    \hline
    num\_a1            & num\_a2     & \gr{num\_a3}  & num\_a4   \\
    so4\_a1            & so4\_a2     & so4\_a3       &            & H2SO4 \\ 
    soa\_a1            & soa\_a2     & soa\_a3       &            & SOAG  \\
    bc\_a1             &             & \gr{bc\_a3}   & bc\_a4    \\
    pom\_a1            &             & \gr{pom\_a3}  & pom\_a4   \\
    ncl\_a1            & ncl\_a2     & \gr{ncl\_a3}  &           \\
    mom\_a1            & mom\_a2     & \gr{mon\_a3}  & mom\_a4   \\
    \gr{dst\_a1}       &             & \gr{dst\_a3}  &           \\
    \hline
    \end{tabular}
    }
    \label{tab:tracers}
\end{table}

\begin{table}[t!] 
    \centering
    \caption{Input variables in addition to those shown in black in Table~\ref{tab:tracers}, which are used in the calculation of aerosol microphysics under cloud-free conditions in E3SMv2.} 
    \vspace{1pt}
    \renewcommand{\arraystretch}{1.2} 
    \scalebox{0.9}{
    \begin{tabular}{lp{8cm}p{6.5cm}}
    \hline
    {\bf Index} & {\bf Physical quantity} & {\bf Note} \\\hline
    1 & Air temperature & - \\
    2 & Air pressure    & - \\
    3 & Pressure layer thickness & Used for diagnosing air densitiy. \\
    4 & Specific humidity & Used for diagnosing relative humidity. \\
    5 & Planetary boundary layer height &  \\
    6 & Geopotential height & Used for comparing with the planetary boundary layer height to determine whether boundary layer nucleation should be calculated.\\
    7 & Pre-chemistry mixing ratio of H2SO4 gas & Used for comparing with the pre-microphysics mixing ratio to derive the chemical production rate used as a forcing term of the ODEs. \\
    8  & Geometric mean dry diameter of accumulation mode & - \\
    9  & Geometric mean dry diameter of Aitken mode & - \\
    10 & Geometric mean dry diameter of coarse mode & - \\
    11 & Geometric mean dry diameter of primary carbon mode & - \\
    12 & Geometric mean wet diameter of accumulation mode & - \\
    13 & Geometric mean wet diameter of Aitken mode & - \\
    14 & Geometric mean wet diameter of coarse mode & - \\
    16 & Geometric mean wet diameter of primary carbon mode & - \\    
    16 & Wet density of accumulation mode particles & - \\
    17 & Wet density of Aitken mode particles & - \\
    18 & Wet density of coarse mode particles & - \\
    19 & Wet density of primary carbon mode particles & - \\
    \hline 
    \end{tabular}
    }
    \label{tab:other_input}
\end{table}

It is worth noting that MAM was originally developed for simulating the troposphere, i.e., the bottom 10--20 km of the atmosphere above the Earth's surface, while the aerosol optical properties in the stratospheric have been prescribed up until v2 of E3SM's standard releases. Consequently, the aerosol mixing ratio changes in the stratosphere calculated by microphysics and other parameterizations are either very close to zero or not consistent with real-world observations. In parallel to our emulation-focused study, a new version of MAM was incorporated into the PDE-ODE system of E3SM for its version 3 release \cite{xie2025eamv3}, with a fifth lognormal mode introduced for stratospheric sulfate aerosols originating from explosive volcanic eruptions. Awareness of that development led to our decision to exclude the top 26 layers of E3SMv2's vertical grid in our exploration of emulation, which correspond to air pressure values of approximately 0.1~hPa to 100~hPa.

\subsection{Machine learning emulation strategy} \label{sub:ml}
Given the intention of potentially replacing the aerosol microphysics parameterization in E3SMv2 by a machine-learned emulator and considering the split nature of the multi-process time integration in the original PDE-ODE system of E3SMv2, we define our emulation task as learning the one-timestep evolution of the 20 aerosol and gas mixing ratios that are affected by the ODEs describing aerosol microphysics in cloud-free conditions (Table~\ref{tab:tracers}). Key design features of the emulators and the training procedure are summarized below.

\subsubsection{Input and output} \label{subsub:inout}
Because the changes of the mixing ratios over a single timestep (30 min) are typically small compared to the starting values and the spatial variability across the globe and at different altitudes,
we define the outputs of the ML emulators as the one-timestep changes instead of the ending values. 
The input features and output variables of the emulators are consistent with the input and output variables of the aerosol microphysics parameterization summarized in Sect.~\ref{sub:amicphys}, which include initial values of 20 mixing ratios (Table~\ref{tab:tracers}) plus the atmospheric conditions, aerosol properties, and chemical production rate of H2SO4 gas listed in Table~\ref{tab:other_input}. 
We note that while the E3SM code takes the planetary boundary layer height (PBLH) and the altitude of grid cell (i.e., geopotential height denoted by zm in the code) as two separate variables, deeper in the code structure, these two quantities are compared to determine whether the grid cell is located within the boundary layer and hence boundary layer nucleation should be calculated. For the emulation task, we combine the two quantities (PBLH and zm) into one binary input feature as a data preprocessing step.

The input and output variables of the microphysics parameterization have different units and vastly different typical value ranges. Therefore, we follow a common practice of transforming the data samples so they have similar statistics. For instance, it is known that neural networks with standard initialization schemes and activations behave best with $O(1)$ values, and their performance may suffer due to numerical artifacts in the presence of disparate scales. $z$-score normalization transforms variables to have a mean of zero and standard deviation of 1, as follows:
\begin{equation}
    \tilde{\phi} = \dfrac{\phi-\mu_{\phi}}{\sigma_{\phi}} \,. \label{eq:norm}
\end{equation}
Here, $\mu_{\phi}$ and $\sigma_{\phi}$ are the mean and standard deviation of the variable $\phi$ while $\tilde{\phi}$ denotes the variable after standardization.
Each input and output variable is normalized independently.

Furthermore, aerosol and gas mixing ratios across different geographical locations and altitudes often span multiple orders of magnitude, with statistical distributions characterized by many small values and a small fraction of large values--both of which are important to capture accurately. To address this multiscale and skewness challenge,  we seek a transformation such that the transformed data ranges across fewer scales. It should be noted that this transformation should be invertible so it can be used during inference, too. While aerosol modelers often use the logarithmic transformation in model equations and analysis, this transformation severely minimizes the variation at the largest extreme values and cannot directly handle negative values. An alternative and numerically appealing transformation is a power-based variable transformation in the form of
\begin{equation}
    \mathcal{T}(r; a) = r^a \quad\text{with}\quad 0 < a < 1\,. \label{eq:power}
\end{equation}
where $r$ is an input mixing ratio or the mixing ratio change caused by aerosol microphysics. Since changes in mixing ratios can span both positive and negative values, we define $a := \dfrac{1}{n}$, where $n$ is an odd number. We note that in the case of nonlinear (power) transformation, we first apply the power transformation (Eq.~\ref{eq:power}), followed by a z-normalization (Eq.~\ref{eq:norm}) as follows:
\begin{equation}
\begin{aligned}
    \phi_a &= \phi^a , \\
    \tilde{\phi} &= \dfrac{\phi_a-\mu_{\phi_a}}{\sigma_{\phi_a}} \,. \label{eq:norm}
\end{aligned}
\end{equation}

\subsubsection{ML architecture} \label{subsub:arch}
As explained in Section~\ref{sec:intro}, our primary goal for this study is to identify key challenges, bottlenecks, and opportunities in emulating aerosol microphysics with ML. Accordingly, we adopt the simple and widely-used FNN as in \cite{harder2021emulating} and \cite{harder2022physics} using the rectified linear unit (i.e., ReLU) as the activation function, deferring the exploration and detailed assessment of advanced architectures to future work. By focusing on a simple architecture, we provide a easy-to-build baseline for future efforts, enabling systematic exploration of design choices and their impact on emulator performance.

A single network is used to emulate the $20$ target variables (i.e., the size of the input layer is 39 and the size of the output layer is 20).
The only architectural modification compared to \cite{harder2021emulating} and \cite{harder2022physics} is the inclusion of residual connections between consecutive layers.
Residual connections introduce explicit identity mappings between layers, such that each layer learns a correction to its input rather than a full nonlinear transformation \cite{he2016deep}. During training, this structure modifies the layer-wise Jacobians from products of nonlinear operators to perturbations of the identity, which improves the conditioning of the resulting optimization problem \cite{he2016identity}. As a consequence, gradient magnitudes are preserved more effectively across depth, reducing both vanishing and exploding gradients and allowing stable training over a substantially wider range of learning rates. This allows us to reliably assess the performance of the emulator across different widths and depths as shown in Section~\ref{sub:complexity} with minimal tuning of the optimizer. We also note that residual connections have been adopted in multiple applications of neural networks in Earth systems science \cite{han2020moist,rasp2021data,wang2022deep}, and they demonstrated superior performance compared to plain networks that do not include these connections.

\subsubsection{Emulator training}

For emulator training, we use a loss function of mean squared error (MSE), defined as follows:
\begin{equation}\label{eq:loss}
    J(\theta) = \dfrac{1}{B}\sum_{i=1}^{B} \sum_{j=1}^{N_\text{var}} \left(\tilde{\phi}_{i,j} - \tilde{\phi}_{i,j}^{\theta}\right)^2,
\end{equation}
where $\tilde{\phi}_{i,j}$ denotes the $i^{th}$ sample of the $j^{th}$ output variable (one-timestep change in mixing ratio), $\tilde{\phi}_{i,j}^{\theta}$ is the corresponding prediction from the emulator, and $N_\text{var}$ is the number of output variables (20 in the current study). We note that both $\tilde{\phi}_{i,j}$ and $\tilde{\phi}_{i,j}^{\theta}$ have been transformed according to Sec.~\ref{subsub:inout}. $\theta$ represents the neural network parameters (i.e., weights and biases), and $B$ denotes the batch size, set as $4096$ in the present study. Our experiments demonstrated that results were qualitatively unchanged for relatively smaller batch sizes (e.g., 256), although these resulted in more iterations per epoch and under-utilization of GPU memory.

For the optimizer, the Adaptive Nesterov momentum algorithm (Adan) is used with a learning rate of $3\times 10^{-4}$. Adan builds on the classic Adam optimizer by incorporating Nesterov momentum for potentially faster convergence and more desitrable performance over longer training runs. Training is progressed for 5000 epochs to allow the optimizer converge to a stable minimum with no early stopping. A discussion on training convergence can be found in Sect.~\ref{sub:convergence}.

\subsection{Training data from an E3SMv2 control simulation} \label{sub:data}
To provide data for training emulators and assessing their accuracy, we performed a simulation with E3SMv2 in a configuration that is commonly used for the development of the atmosphere component, namely the ``F2010 component set'', which involves active atmosphere and land components.
The sea surface temperature and sea ice extent, as well as the external sources of aerosols and precursors, were specified using multi-year mean of 2005–2014 with repeating annual cycles.
Due to this setting, the years and dates in the simulation are not expected to correspond to actual dates in the recent history, although the simulated atmospheric states in different months are expected to represent characteristics of the corresponding seasons under the average conditions of 2005-2014. 

The computational mesh used by the atmospheric physics parameterizations is the ne30pg2 cubed sphere grid with a total of $21600$ grid columns \cite{hannah2020grids}, corresponding to horizontal grid spacings of about 165~km.
The vertical grid had 72 layers extending from the Earth's surface to about 0.1~hPa \cite{rasch2019eamv1,xie2018understanding}.
Time integration was performed using E3SMv2's default setting for this spatial resolution, in which 30~min timesteps were used by aerosol microphysics and many other parameterizations.

The simulation was started on September 1 using initial conditions obtained from a previously performed simulation, and the first 4 months were considered a spin-up phase.
Instantaneous values of input and output variables of the aerosol microphysics parameterization suite were archived for emulator training and testing. The results shown in Section~\ref{sec:ml-results} correspond to daily values from 00~UTC on January 1 through 10. We note that this choice of time slices is somewhat arbitrary; it reflects our intention to focus on a single season in the first step of our exploration, leaving the topic of sampling strategy and emulator generalizability across seasons and forcing scenarios to subsequent investigations.

E3SMv2's aerosol microphysics parameterizations are based on ODEs (i.e., the equations do not involve spatial derivatives), hence we consider each grid cell and time slice as a separate data sample. Due to the decision to emulate microphysical processes affecting interstitial aerosols under cloud-free conditions in the troposphere (cf. Section~\ref{sub:amicphys}), all data points in the top 26 layers of the vertical grid and all data points with cloud presence (cloud fraction $> 0$) are excluded.
This yields over $6.6 \times 10^6$ samples, which we randomly split into training, validation, and testing sets in proportions of $50\%$, $25\%$, and $25\%$, respectively. 
Training samples are used to optimize neural network weights, validation samples are used to guide hyperparameter selection, and testing samples are used to evaluate the selected emulators.

\section{Results} \label{sec:ml-results}
We use the dataset generated by E3SM as described in Section~\ref{sub:data}, and adopt the preprocessing operations outlined in Section~\ref{subsub:inout}. In particular, we apply the power transformation (with $a=\dfrac{1}{3}$), followed by $z$-score transformation. Other choices of exponent for the power transformation were also explored, and the results will be reported separately.  We explore key design choices that influence the performance of the ML emulator designed in Section~\ref{subsub:arch}. In particular, Section~\ref{sub:convergence} illustrates how training convergence is monitored to ensure that the optimizer, learning rate, and batch size are appropriately configured. In Section~\ref{sub:complexity}, we present performance metrics and examines how network depth and width affect accuracy, guiding the selection of the final architecture. 

\subsection{Training procedure and convergence} \label{sub:convergence}
We first assess training convergence to verify that the chosen optimizer (Adan), learning rate ($3\times10^{-4}$), and batch size ($4096$) are appropriate for the emulation task. To this end, we present the suboptimality gap, namely the absolute difference between the loss value at a given epoch and the minimum loss achieved over the entire training period. This metric provides insight into both the final training performance and the optimizer’s progression toward it.

\begin{figure}[ht!]
    \centering
    \includegraphics[width=\linewidth]{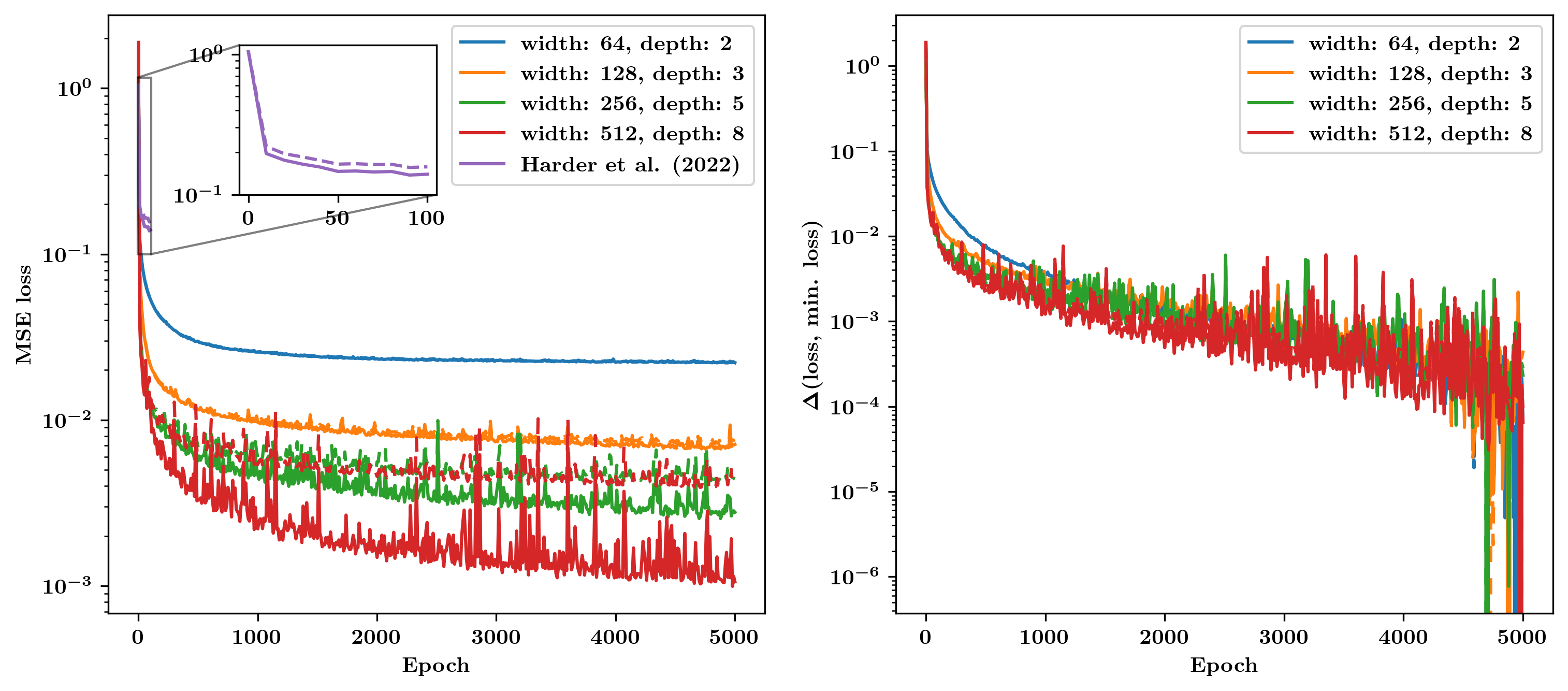}
    \caption{Evolution of the MSE loss (left) and the suboptimality gap (right), namely, the difference between the loss at a specific epoch and the lowest loss observed during optimization, in a subset of offline training experiments. Solid lines refer to the training loss while dashed lines correspond to validation loss. For reference, performance using the framework in \cite{harder2022physics} is shown in purple.
    }
    \label{fig:conv}
\end{figure}

Figure~\ref{fig:conv} shows the history of the training and validation losses (left panel) as well as the evolution of the suboptimality gap (right panel) over 5000 training epochs for four networks with different depths and widths. In all cases, the gap decreases monotonically by more than three orders of magnitude within the first $1000$ epochs and subsequently plateaus at approximately $10^{-3}$--$10^{-4}$. A smaller learning rate slows down the initial decay, whereas a larger learning rate could induce oscillatory behavior as the optimizer overshoots minima in the loss landscape. Together, these results indicate that the selected training configuration yields reliable and well-behaved convergence across network architectures. For comparison, we also shown in the left panel the training and validation loss histories using the configuration from \cite{harder2022physics} applied to our data, which clearly shows that the optimizer requires a sufficient number of epochs to achieve proper convergence; otherwise, satisfactory performance might not be attained. In other words, careful selection of the optimizer, batch size, and learning rate is essential, along with allowing sufficient training time for the optimizer to converge.

\subsection{Emulator's architecture complexity} \label{sub:complexity}
We investigate how architectural complexity influences emulator performance through a systematic grid search over network depth and width. Specifically, the number of hidden layers is varied over $\{1, 2, 3, 5, 8\}$, while the number of neurons per hidden layer is chosen from $\{32, 64, 128, 256, 512\}$. The input and output layer sizes are fixed at $39$ and $20$, respectively, consistent with the dimensionality of the emulator’s input and output vectors. All emulators are trained using identical datasets and optimization settings, and their performance is evaluated on the validation set. For comparison across architectures, we consider two complementary performance metrics: the coefficient of determination ($R^2$), 
and the relative root mean squared error ($\text{Relative RMSE}$), defined as:

\begin{align}
    R^2\ (j) &= 1 - \dfrac{\sum\limits_{i \in \mathcal{I}_{valid}}(y_{i,j}^{true} - y_{i,j}^{NN})^2}{\sum\limits_{i \in \mathcal{I}_{valid}}(y_{i,j}^{true}-\bar{y}_j)^2}, \\
    \text{Relative RMSE}\ (j) &= \dfrac{\sqrt{\sum \limits_{i \in \mathcal{I}_{valid}} (y_{i,j}^{true} - y_{i,j}^{NN})^2}}{\sqrt{\sum \limits_{i \in \mathcal{I}_{valid}} (y_{i,j}^{true})^2} }
\end{align}
where $y_{i,j}^{true}$ denotes the E3SM-computed $30$-minute change of the $j^{th}$ mixing ratio for sample $i$, $y_{i,j}^{NN}$ is the corresponding emulator prediction, and $\mathcal{I}_{valid}$ is the index set for the validation dataset. We compute these metrics for each mixing ratio individually, as well as their averaged values. In addition, we define a norm-based scores (using transformed variables) as follows:
\begin{align}
    R^2\ (norm) &= 1 - \dfrac{\sum\limits_{i \in \mathcal{I}_{valid}} \sum\limits_{j=1}^{20} (\tilde{y}_{i,j}^{true} - \tilde{y}_{i,j}^{NN})^2}{\sum\limits_{i \in \mathcal{I}_{valid}} \sum\limits_{j=1}^{20}(\tilde{y}_{i,j}^{true}-\bar{\tilde{y}}_j)^2}, \\
    \text{Relative RMSE}\ (norm) &= \dfrac{\sqrt{\sum \limits_{i \in \mathcal{I}_{valid}} \sum\limits_{j=1}^{20} (\tilde{y}_{i,j}^{true} - \tilde{y}_{i,j}^{NN})^2}}{\sqrt{\sum \limits_{i \in \mathcal{I}_{valid}} \sum\limits_{j=1}^{20} (\tilde{y}_{i,j}^{true})^2} }
\end{align} \label{eq:norm-score}

\begin{figure}[ht!]
    \centering
    \includegraphics[width=\linewidth]{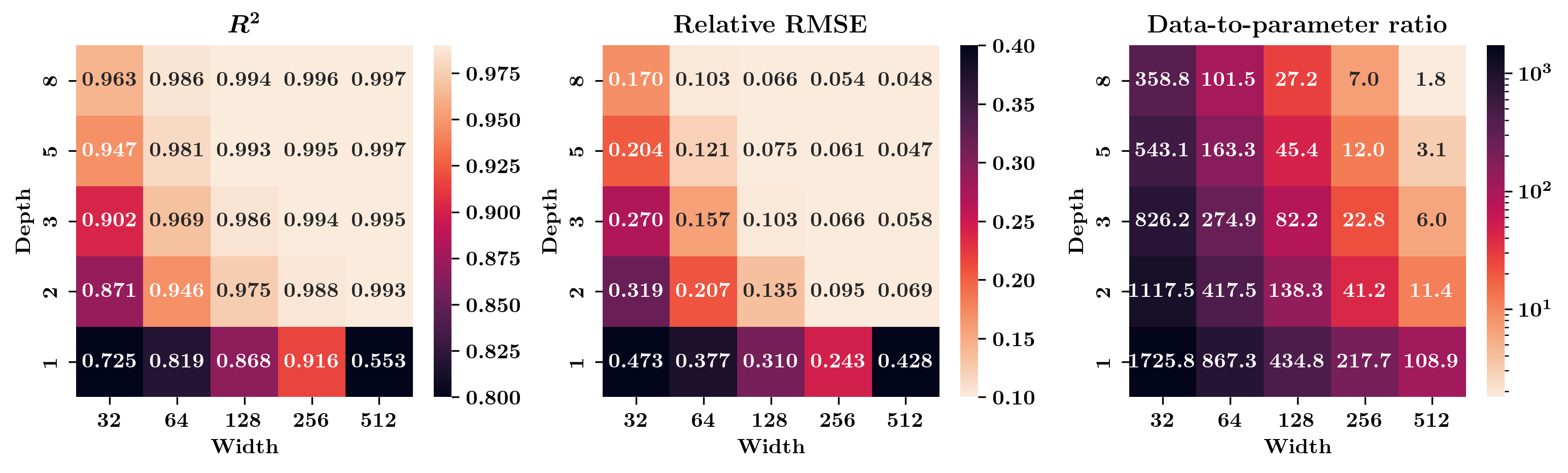}
    \caption{Heatmaps of the performance metrics (at validation dataset) for different emulators with varying widths and depths, averaged across the 20 target variables.}
    \label{fig:heatmap_width_depth}
\end{figure}

\begin{figure}[ht!]
    \centering
    \includegraphics[width=\linewidth]{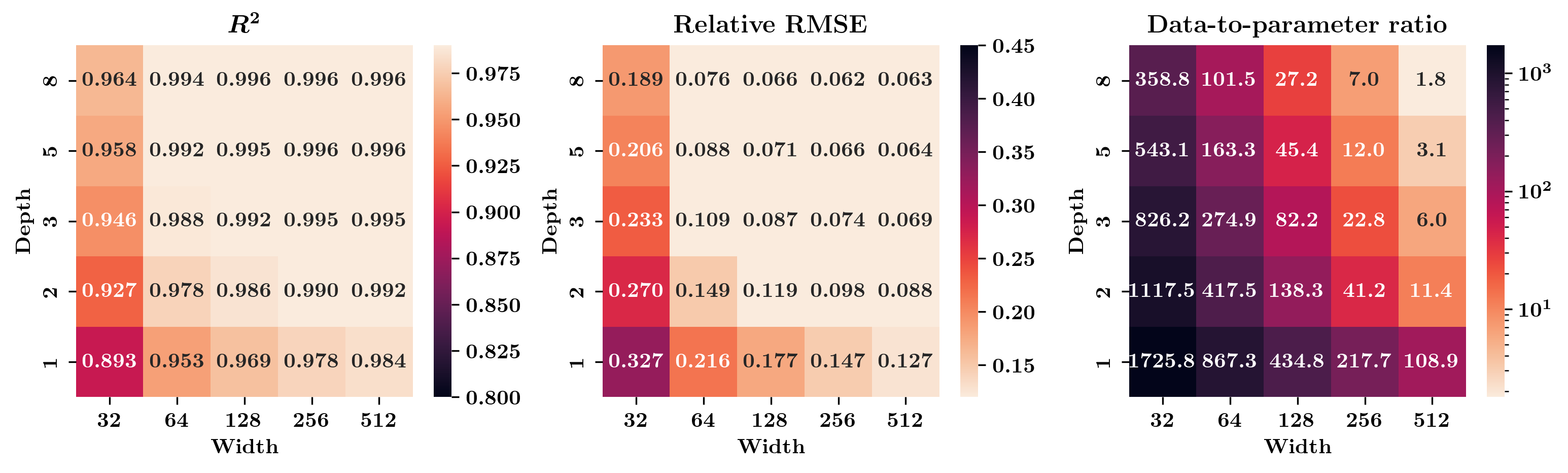}
    \caption{Heatmaps of the \textit{norm-based} performance metrics (at validation dataset) for different emulators with varying widths and depths.}
    \label{fig:heatmap_width_depth_norm}
\end{figure}

\begin{figure}[ht!]
    \centering
    \includegraphics[width=\linewidth]{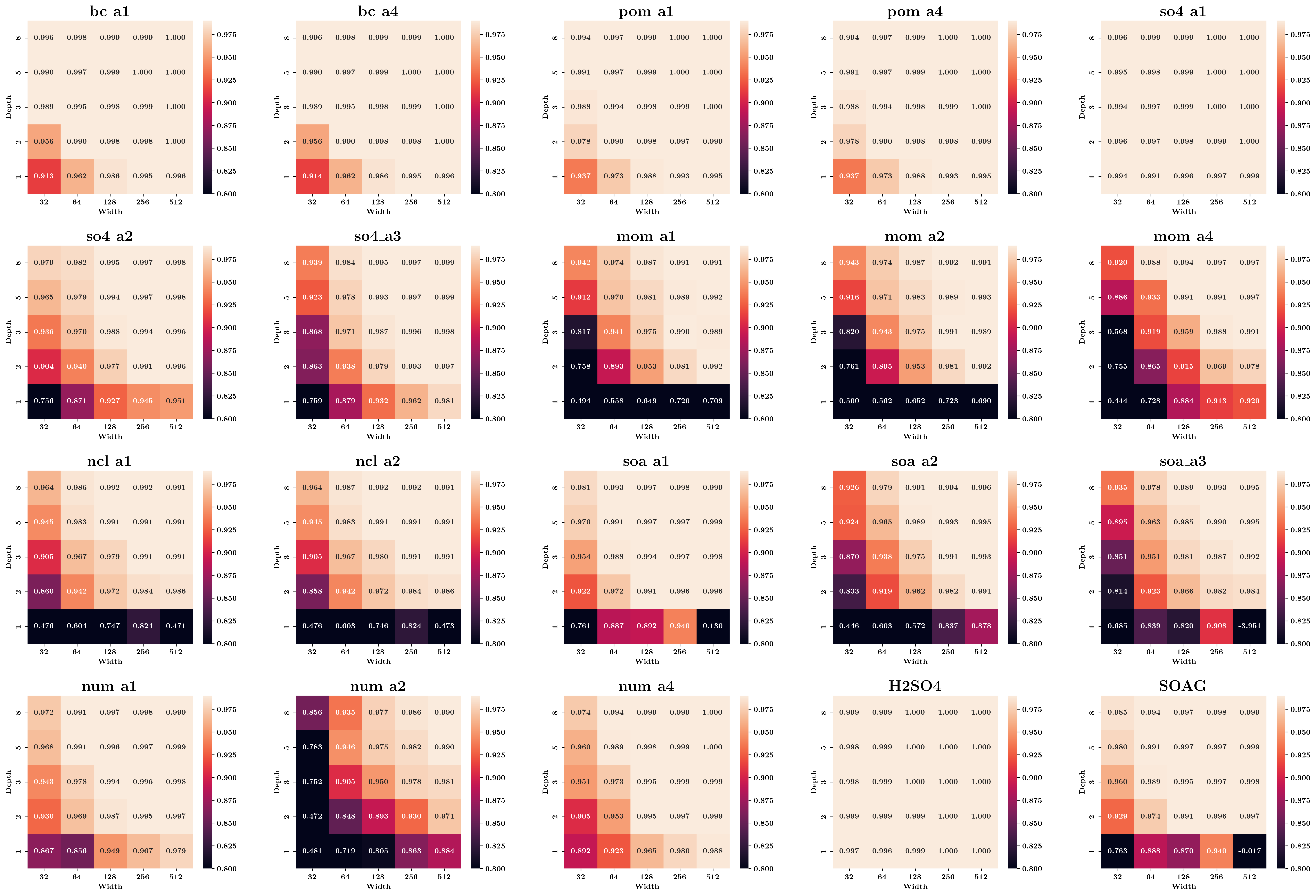}
    \caption{Heatmaps of the $R^2$ (at validation  dataset) for each target variable with varying widths and depths.}
    \label{fig:heatmap_width_depth_variable}
\end{figure}

Figure~\ref{fig:heatmap_width_depth} summarizes the results through heatmaps of the performance metrics (averaged across the 20 target variables) for different width–depth combinations while Figure \ref{fig:heatmap_width_depth_norm} shows the norm-based scores in Eq.~\ref{eq:norm-score}. In addition, Figure~\ref{fig:heatmap_width_depth_variable} depicts the variation of validation $R^2$ values for each of the target variables as a function of model's width and depth. Several clear trends emerge:

First, the shallowest and narrowest networks, i.e. those with only 1 hidden layer with 32 to 512 neurons per layer, or with 1 to 8 hidden layers but only 32 neurons per layer, feature the lowest accuracies, indicating that extremely shallow or narrow networks are insufficient to capture the complexity of aerosol microphysics.

Second, while increasing the number of hidden layers from 1 to 2 leads to substantially better accuracy across all network widths explored here, further increases in depth become less beneficial than increases in width. One possible interpretation is that a wider network potentially learns a more meaningful latent space with distinct clusters in the dataset, while the forward mapping may not be complex enough to necessitate a significantly deeper network (beyond $2$ or $3$ hidden layers). In this study, we select the $3\times256$ architecture as a principled compromise between predictive accuracy, architectural parsimony, and computational efficiency. All subsequent results are reported using this configuration.

Third, a clear stratification emerges across variables: some mixing ratio changes are consistently emulated with high fidelity, while others exhibit substantially lower $R^2$ scores. Notably, the latter groups includes aerosols originating from the ocean, namely the marine organic matter (mom) and sea salt (ncl), mixing ratios related to secondary organic aerosols (soa and SOAG), as well as the Aitken mode particle number mixing ratio (num\_a2). This distinction motivates a potential extension of the present work in which multiple emulators are employed to different variables, with more complex architectures reserved for challenging variables, while simpler networks provide sufficient accuracy for others. This approach may lead to more efficient resource allocation and more uniform prediction accuracy across variables.

\clearpage
\subsection{Emulator performance across target variables} \label{sub:results}
\begin{figure}[ht!]
    \centering
    \includegraphics[width=\linewidth]{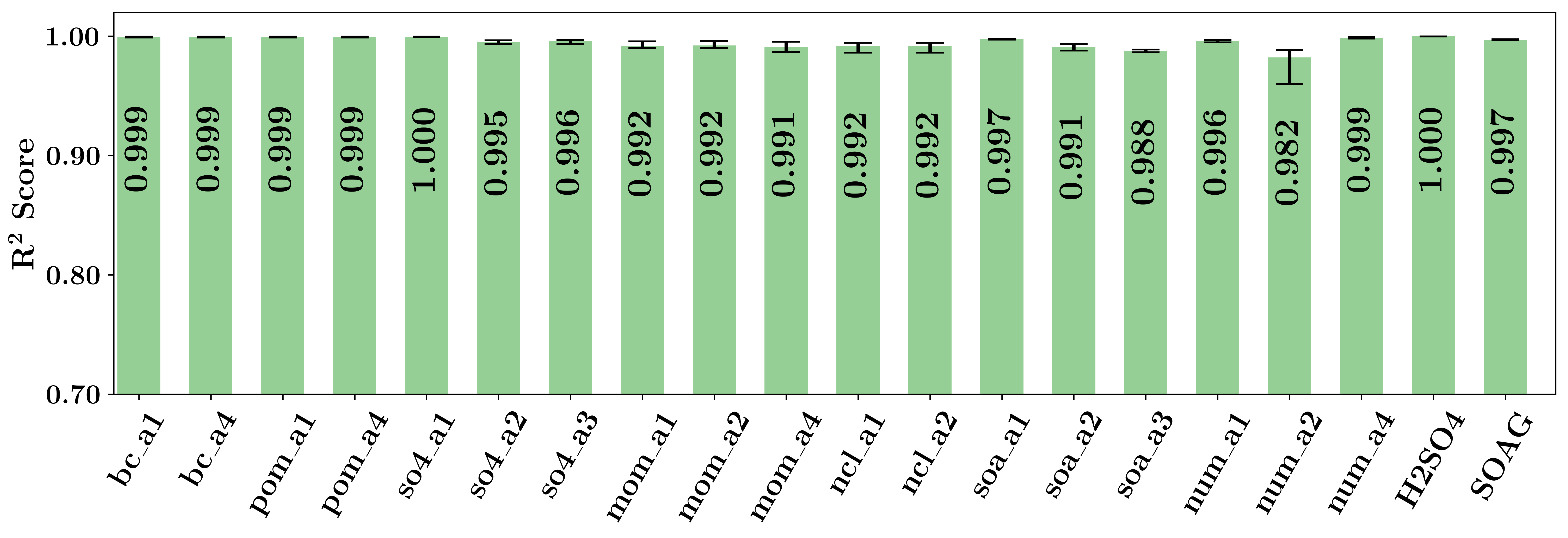}
    \caption{$R^2$ values of ML emulator for each of the $20$ target variables. Bar height corresponds to the average daily score across from January 1 to January 10, and error bars designate the minimum and maximum scores obtained in these 10 days.}
    \label{fig:r2}
\end{figure}

In this section, we evaluate the emulation performance of the selected architecture, consisting of three hidden layers with $256$ neurons per layer. 
Figure~\ref{fig:r2} reports the coefficient of determination ($R^2$) for each of the 20 emulated variables, computed on the testing dataset. We observe that the emulator achieves around ~0.99 scores across all variables, using a relatively simple architecture. We also indicate that the power transformation in Sec.~\ref{subsub:inout} has enabled this satisfactory performance across different targets, spanning a disparate range of magnitudes. In an extended manuscript that we plan to share later, we will further show how the performance changes if we stick to linear transformations (e.g., z-score normalization) as well as the effect of different exponents in the power transformation.

\begin{figure}[ht!]
    \centering
    \includegraphics[width=\linewidth]{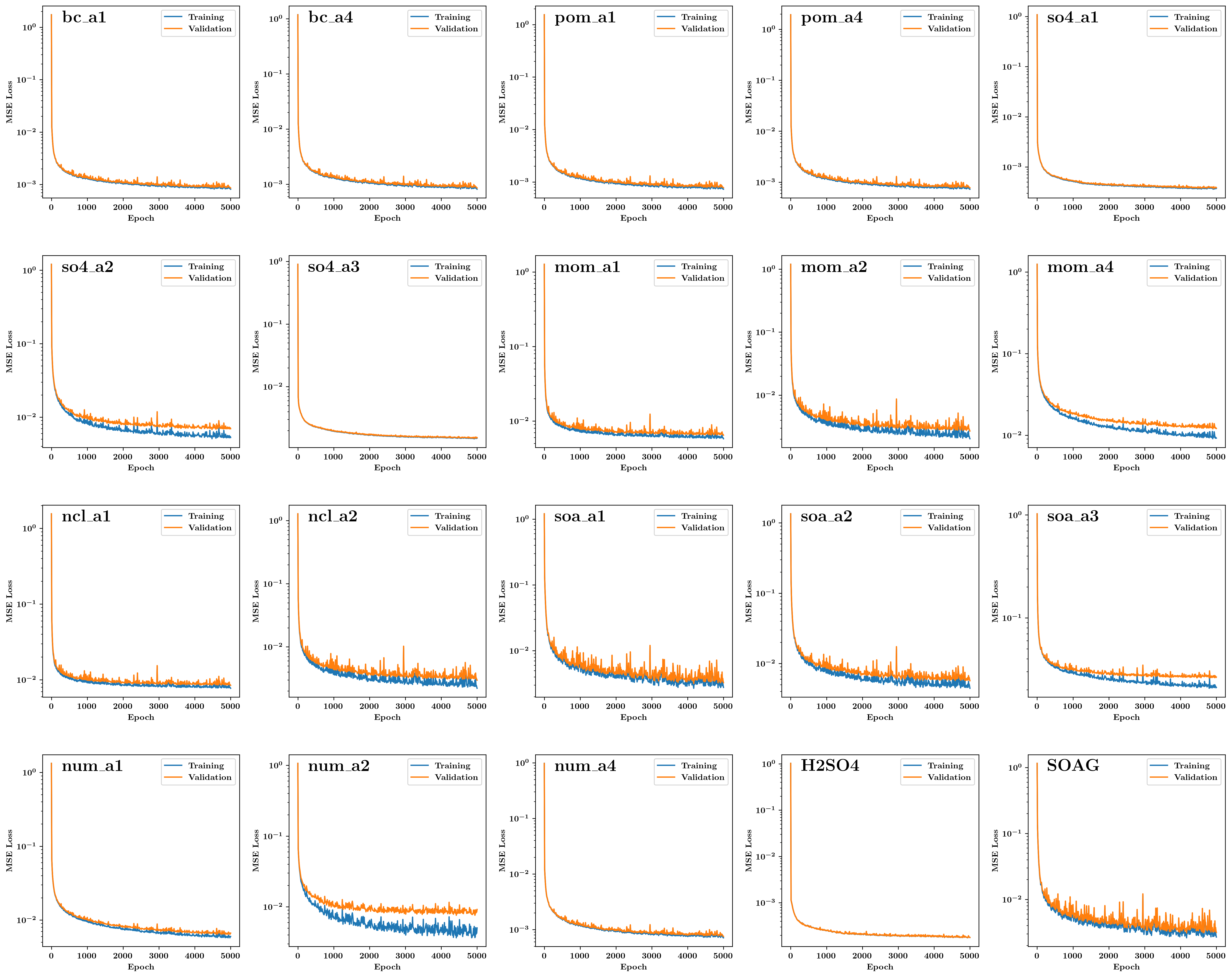}
    \caption{History of training and validation losses for each of the $20$ emulated variables, using a feedforward neural network with $3$ hidden layers, each having $256$ neurons.}
    \label{fig:train_valid_history}
\end{figure}

\begin{figure}[ht!]
    \centering
    \includegraphics[width=\linewidth]{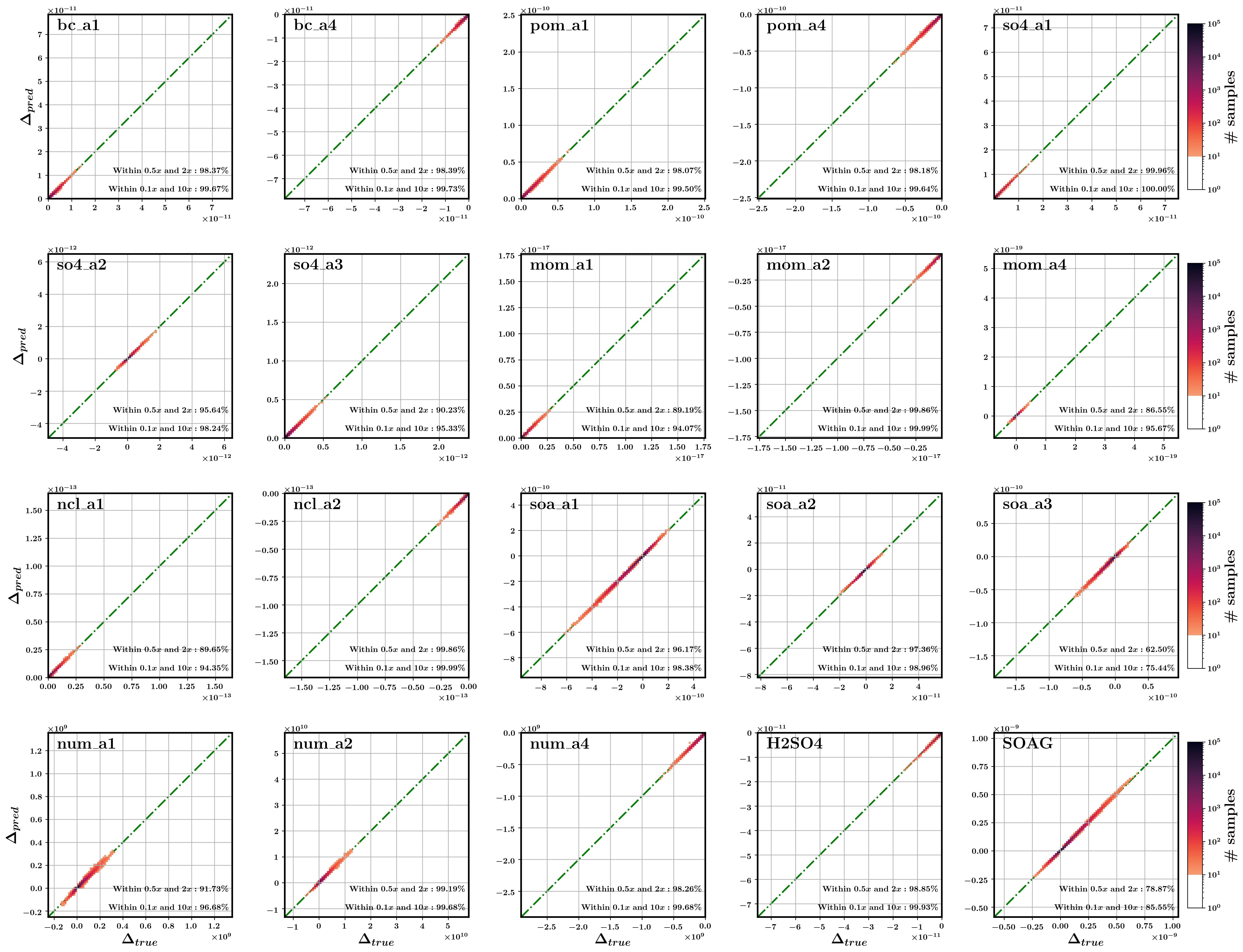}
    \caption{2D histograms comparing the one-timestep mixing ratio changes in the testing dataset predicted by E3SMv2 (x-axis) and by an ML emulator with 3 hidden layers and 256 neurons per layer.}
    \label{fig:hist_true_vs_pred}
\end{figure}

Another important perspective for assessing the capabilities and limitations of ML-based emulation is to examine how predictive accuracy varies across value ranges. Figure~\ref{fig:hist_true_vs_pred} presents two-dimensional histograms of true versus predicted $30$-minute changes in mixing ratios for selected variables. The color denotes the number of testing samples falling within each magnitude range. Overall, the majority of samples are concentrated along the one-to-one line, indicating strong agreement between predicted and true values across much of the data distribution. The figure also shows the percentage of predictions that fall within the 1:2 and 2:1 ratios with respect to their true values, which provide another indication of the performance and coverage of the emulator. We note that much of these desirable results can be attributed to the nonlinear transformation defined in Sec.~\ref{subsub:inout}. Otherwise, a simple linear transformation will provide good results for values of the largest magnitudes while suffering in regions corresponding to smaller-magnitude changes.
In addition, long sessions of training were also found to be important, so that we could avoid relatively large (but still reducible) training and validation errors caused by early stopping, and thereby help each emulator configuration reach its full potential. 
We will provide more detailed results in these aspects in the follow-up extended manuscript.

\section{Concluding Remarks} \label{sec:conclusion}
We explored machine learning (ML) emulation of the aerosol microphysics parameterization suite for cloud-free conditions in E3SMv2, to assess to which extent a neural network-based emulator could accurately reproduce the behavior of the current microphysics paramterization. By focusing first on emulating the current operational parameterization, we established a controlled benchmark problem that isolates emulator fidelity before pursuing more ambitious ML replacements based on process-resolved models or observational constraints. Our results show that feedforward neural networks (FNNs) provide a viable framework for emulating aerosol microphysics, but emulator performance depends strongly on architecture design, training convergence, and the treatment of values that differ by many orders of magnitude. In particular, the use of appropriate variable transformations (e.g., power transformation), careful choices of hyperparameters that ensure training convergence, and careful evaluation of model complexity are essential for obtaining balanced skill across target variables and across different ranges of characteristic values of the targets. These findings underscore that emulator accuracy should not be judged from a single aggregated metric; performance must be examined across the full set of predicted variables and various metrics should be considered. These considerations are especially relevant for multiscale problems with stiff, nonlinear, or highly heterogeneous dynamics.
Beyond this specific application of aerosol microphysics emulation, we believe several aspects of our conclusions are generally applicable to Earth system model emulation. Strategies such as monitoring convergence behavior, exploring architecture complexity systematically, and transforming variables to improve learnability in different value ranges are likely relevant not only to aerosol microphysics, but also to other parameterizations in large-scale models as well other PDE- and ODE-based process models.

Our work so far focused only on a single month of the year and assessed emulator accuracy in offline testing using data archived from E3SM simulations. Two natural next steps are to include full seasonal cycles and to assess online performance within the host model E3SM, including numerical stability and end-to-end impacts on multi-year statistics of the simulated aerosol life cycles and atmospheric conditions.

\section*{Acknowledgments}
This work was supported primarily by the U.S. Department of Energy, Office of Science, Scientific Discovery through Advanced Computing (SciDAC) program, via a partnership in Earth System Model Development between the Advanced Scientific Computing Research (ASCR) and the Biological and Environmental Research (BER), as part of the project titled ``Physical, Accurate, and Efficient atmosphere and surface coupling across SCALes'' (PAESCAL; proposal No. 0000267817). 
PS was supported by the LEarning-Accelerated Domain Science (LEADS) SciDAC Institute (Project No. 85462) to provide guidance on emulator design, training and evaluation.
KZ was supported by the Energy Exascale Earth System Model (E3SM) project to provide information on the aerosol microphysics parameterization suite in E3SMv2. 
All other authors were supported by the PAESCAL SciDAC partnership project.
The work used resources of the National Energy Research Scientific Computing Center (NERSC), a U.S. Department of Energy Office of Science User Facility located at Lawrence Berkeley National Laboratory, operated under Contract Number DE-AC02-05CH11231, using NERSC awards ASCR-ERCAP33440, ASCR-ERCAP28881, and BER-ERCAP30805. Pacific Northwest National Laboratory is operated for the U.S. Department of Energy by Battelle Memorial Institute under contract DE-AC06-76RLO1830.

\bibliographystyle{unsrtnat}
\bibliography{references}

\end{document}